\documentclass[letterpaper,aps,superscriptaddress,showpacs,floatfix,twocolumn]{revtex4-1}

\usepackage{hyperref}
\usepackage{color}
\usepackage{graphicx}	% Include figure filed
\graphicspath{%
  {./},%
}

\usepackage{xspace}	% Include xspace

\begin{document}

\title{Extracting the shear viscosity of a high temperature hadron gas}

\author{Paul Romatschke}
\affiliation{Department of Physics, 390 UCB, University of Colorado, Boulder, 80309-0390, USA}
\author{Scott Pratt}
\affiliation{Department of Physics and Astronomy and National Superconducting Cyclotron Laboratory,
Michigan State University, East Lansing, Michigan 48824, USA}

\date{\today}

\begin{abstract}

Quark-Gluon plasmas produced in relativistic heavy-ion collisions quickly expand and cool, entering a phase consisting of multiple interacting hadronic resonances just below the QCD deconfinement temperature. The transport properties of this hot hadron gas are poorly understood, yet they play an important role in our ability to infer transport properties of the quark-gluon plasma, because experimental measurements integrate over the whole system evolution. Assuming that the hot hadron gas can be modeled by a hadron cascade code based on kinetic theory assuming binary interactions, the shear viscosity over entropy ratio of a hot hadron gas for temperatures in between 120 MeV and 170 MeV is extracted.
%The shear viscosity matches well with simple expressions based on the density and collision rate.
Furthermore, we present estimates for a second order transport coefficient, the shear viscous relaxation time at a temperature of 165 MeV.
\end{abstract}

\maketitle

The theory of strong interactions, QCD, predicts that at temperatures exceeding $T_c\simeq 170$ MeV, ordinary hadrons dissolve into their constituents creating a state of matter called the 'quark-gluon plasma'. From first-principle lattice QCD studies, it is known that the transition from confined quarks and gluons (hadrons) to the deconfined quark-gluon plasma is not a true, sharp phase transition but rather an analytic cross-over transition \cite{Aoki:2006we,Bhattacharya:2014ara}. Relativistic heavy-ion collisions, conducted at different collision energies at the Relativistic Heavy-Ion Collider (RHIC) and the Large Hadron Collider (LHC) aim at understanding the properties of QCD matter by creating quark-gluon plasmas at temperatures up to $\sim 450$ MeV, which then expand and cool to temperatures close to $T_c$ within a time of $\tau_0 \sim 10$ fm/c ($10^{-22}$ seconds). Below $T_c$, quarks and gluons are thought to confine into color-neutral objects (hadrons and hadron resonances) in a process called hadronization which is still not understood in detail. This hadron gas then continues to expand and cool, with the different hadrons scattering, decaying and merging for several tens of fm/c, before the stable remaining particles cease to interact and are eventually recorded by the experimental detectors. With the possible exception of direct photons and dileptons, all experimental information on the transport properties of hot QCD matter must be extracted from these final state hadrons. 

From the creation to the time of last particle interaction, the systems created in heavy-ion collisions spend most of their lifetime in the hadronic gas phase. As such, inferring properties of the quark-gluon plasma phase from the measured particle spectrum with good precision naturally requires good understanding of the properties of the hadron resonance gas. While this is arguably the case at low temperatures where the system is well approximated as a gas of only pions \cite{Gavin:1985ph,Prakash:1993bt,Dobado:2001jf,Chen:2006iga,Chen:2007xe,Itakura:2007mx}, at temperatures close to $T_c$ the properties of this hadron resonance gas are poorly known. Standard modeling procedure in heavy-ion collisions (which will also be used here) employs so-called hadron cascade codes to describe the hot hadron gas, which essentially consist of a kinetic theory simulation including all the hadronic resonance states found in the particle data book. 
%along with their interaction cross-sections. 
Despite residual uncertainty in this model which arises from  poorly known cross sections in the temperature regime close to $T_c$, this approach is considered the best current available model. 
% that are often extrapolated to the temperature regime below $T_c$. While one might worry about the accuracy of any implementation of this particular description, there seems to be consensus in the community that this nevertheless provides the best current available model. However, even under the assumption that hadron cascade codes provide perfect description of the hot hadron gas in nature, very little is known about its transport properties close to $T_c$.
%Unfortunately, these interaction cross-sections are typically not known at temperatures close to $T_c$, so extrapolation 
However, very little is known about its transport properties close to $T_c$. Two studies in the literature attempted to extract the value of shear viscosity over entropy density $\eta/s$ from a particular hadron cascade code (URQMD, Ref.~\cite{Bass:1998ca}), using different techniques, and finding inconsistent values \cite{Demir:2008tr,Song:2010aq}. 
%Indeed, the second of these studies concluded that the hadron gas $\eta/s$ could not be calculated at all, given that they failed to find single valued results for at given temperature. 
%No study on the bulk viscosity over entropy ratio $\zeta/s$ in a hadron cascade code exists, nor do there seem to exist any studies on the importance and effect of second-order transport coefficients (such as relaxation times in the shear and bulk sector), respectively.
%
%
%As of today, standard procedure in the modeling of heavy-ion collisions is to describe the system at high temperatures using relativistic viscous fluid dynamics (with shear and bulk viscosities as free parameters), stopping the evolution when the system has reached a predefined switching temperature $T_{SW}$, and then following the rest of the evolution using a given hadron cascade code. As such, the final resulting particle spectra depend (significantly) on the choice of the switching temperature $T_{SW}$, which enters as an additional model parameter.
%
The present work is meant to fill a gap in our understanding of the hot hadron gas by providing procedures to extract transport properties from hadron cascades, and discuss findings in relation to our knowledge about transport in hot nuclear matter. The specific goal is to examine the details of the stress-energy tensor to reliably extract effective viscosities. Since experimental measurements in heavy-ion collisions deliver observables that are time-integrated over quark-gluon plasma and hadron gas phase, detailed knowledge of the hadron gas transport properties are necessary when aiming at precision studies of quark-gluon plasma properties. The main aim of this letter is to deliver quantitatively reliable results on one of these transport parameters, the shear viscosity.

\section*{Methodology and Results}

Ideally, one would want to perform extractions of transport coefficients using standard procedures such as Kubo relations, which rely on measurements of energy-momentum tensor correlators close to equilibrium with infinitesimally small velocity gradients. In such definitions the viscosity can be dominated by a single mode of the fluid that equilibrates very slowly, even if only a fraction of the fluid is in that mode. A more operational definition of the viscosity is to consider the the deviation of the stress-energy tensor from equilibrium in response to a velocity gradient. For our case we consider a velocity gradient of the scale encountered in the hadronic stage of a heavy-ion collision, at a time $\tau_0\geq 10$ fm/$c$. This velocity gradient can be controlled by considering a one-dimensional boost-invariant, or Bjorken, expansion, $v_z=z/t$, with translational invariance in the $x-y$ plane \cite{Bjorken:1982qr}. All local intrinsic quantities then depend solely on the ``proper time'' $\tau\equiv\sqrt{t^2-z^2}$. In viscous hydrodynamics, the stress tensor in the Landau-Lifshitz frame may then be decomposed uniquely as 
\begin{equation}
T^{ab}=\epsilon u^a u^b - (P-\Pi) \Delta^{ab}+\pi^{ab}\,,
\end{equation}
using the local energy density $\epsilon$, pressure $P$, fluid four velocity $u^\mu$, (traceless) shear stress $\pi^{\mu\nu}$ and bulk stress $\Pi$, as well as the (mostly minus signature convention) space-time metric $g_{\mu\nu}$ which is used to define the projector $\Delta^{ab}=g^{ab}-u^a u^b$.
The constitutive relations for arbitrary uncharged fluids have been worked out in full generality to second order in gradients in Ref.~\cite{Baier:2007ix,Romatschke:2009kr} (see also Refs.~\cite{Bhattacharyya:2008jc,Denicol:2012es}) for special cases of weakly and strongly coupled fluids). Rewriting the equations in terms of evolution equations for $\Pi$ and Muronga's $\Phi=-\pi_{zz}$ \cite{Muronga:2003ta} one finds
\begin{eqnarray}
\label{eq:evol}
\Phi&=&\frac{4 \eta}{3\tau}-\eta \tau_\pi
\left[ \partial_\tau \left(\frac{\Phi}{\eta}\right)+\frac{\Pi \Phi}{3\zeta \eta}\right]
-\frac{\lambda_1}{2 \eta^2} \Phi^2-\eta \tau_\pi^* \frac{\Pi \Phi}{3 \zeta \eta}\nonumber\\
\Pi&=&\frac{\zeta}{\tau}-\zeta \tau_\Pi \partial_\tau \left(\frac{\Pi}{\zeta}\right)
-\xi_1 \frac{3 \Phi^2}{2\eta^2}-\xi_2 \frac{\Pi^2}{\zeta^2}\,,
\end{eqnarray}
which are correct up to (including) terms of order ${\cal O}(\tau^{-2})$. Besides the familiar shear and bulk viscosities $\eta,\zeta$, Eqns.~(\ref{eq:evol}) also contain second-order transport coefficients $\tau_\pi,\tau_\Pi,\tau_\pi^*,\lambda_1,\xi_1,\xi_2$ which will affect the dynamical evolution if the system experiences strong gradients or the shear and bulk stresses $\Phi,\Pi$ are not close to their hydrodynamic equilibrium values (the first term on the r.h.s. of Eqns.~(\ref{eq:evol}), respectively.

For this study, the cascade code B3D \cite{Novak:2013bqa} was initialized over a large transverse area so that the properties near $x=y=0$ were unaffected by the boundary, and properties about the stress tensor are extracted at mid-rapidity ($z=0$). The cascade was initialized with particles consistent with thermal and chemical equilibrium, but with the initial momentum distribution modified using the method in \cite{Pratt:2010jt} to adjust the initial anisotropy of the stress-energy tensor.
The shear stress deviation of the stress-energy tensor is given by
\begin{equation}
\Phi=\frac{1}{3}(T_{xx}+T_{yy}+T_{zz})-T_{zz}=\frac{1}{V}\sum_i^{\rm resonances}\frac{p_i^2}{3E_i}- \frac{p_{z,i}^2}{E_i},
\end{equation}
where $V$ is the simulated space volume and the sum is over all hadron resonances in the particle data book with masses up to $2.2$ GeV. 
%$\Phi$ was extracted from B3D as a function of $\tau$ and is shown in Fig. \ref{fig:pizz} for several initial values of $\Phi$. 
%\begin{figure}
%\centerline{\includegraphics[width=0.5\textwidth]{analfit}}
%\caption{\label{fig:pizz}
%The evolution of the stress-energy tensor as extracted from the microscopic sim%ulation are shown as circles, and lines represent solutions to the Israel-Stewart equations with best-fit values for $\chi$ and $\gamma$. This fit suggests a viscosity $\eta/s=0.18$ at $T=165$ MeV. ???? I'll add a legend to the figure that describes how there are 6 lines, one for each initial value of $\pi_{zz}$. I used the same ones you did, i.e. for NS they would correspond to $\eta/s=0.08, 0.16, 0.24, 0.32, 0.4$ and 0.48.???????
%}
%\end{figure}

\begin{figure}[t]
\centerline{\includegraphics[width=0.5\textwidth]{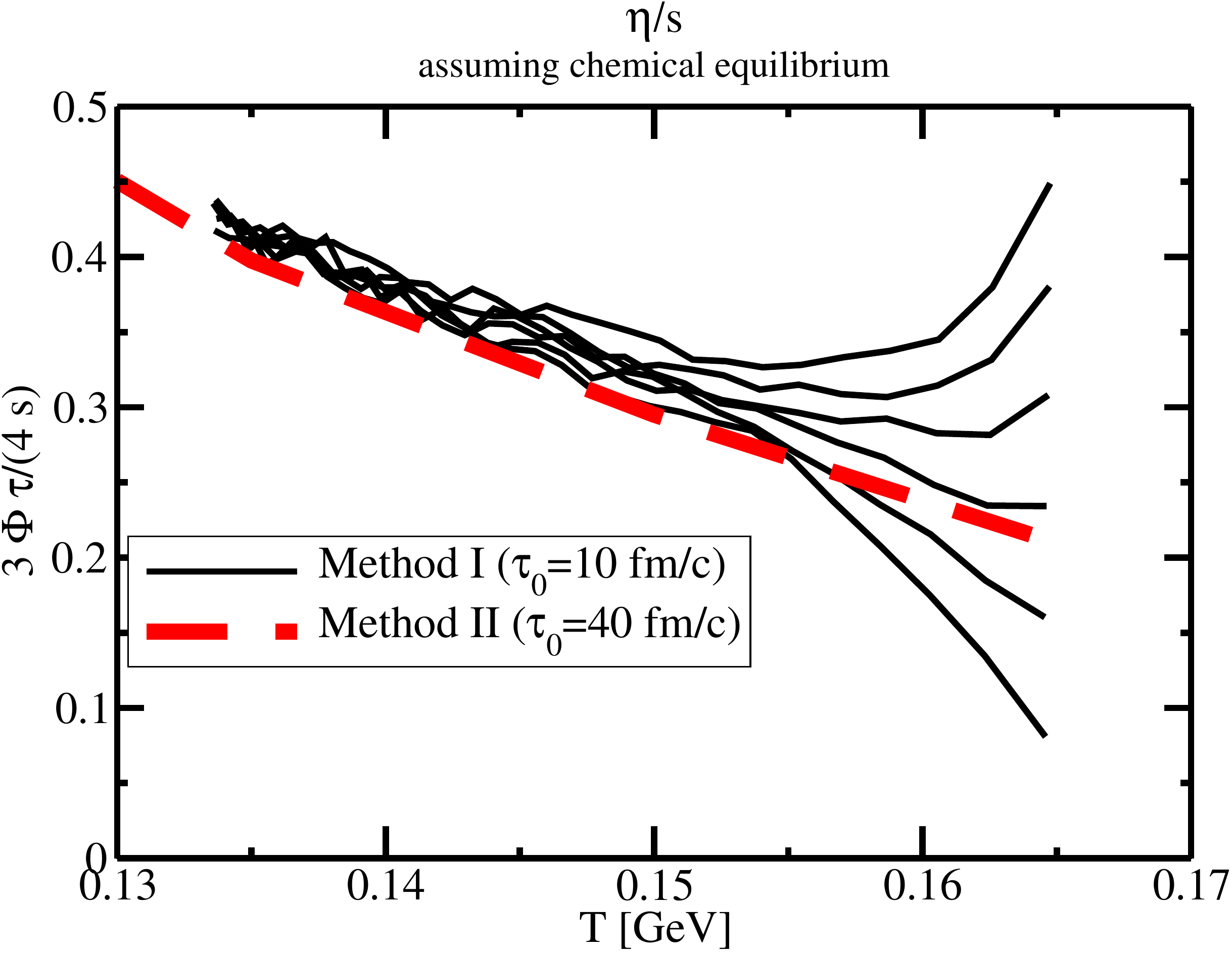}}
\caption{\label{fig:eta}
Shear viscosity over entropy ratio of a hot hadron gas. Shown are two methods to determine $\eta/s$ in the hadron cascade. Method I directly measures $\frac{3 \Phi \tau}{4s}$ whereas Method II determines the value of $\frac{3 \Phi \tau}{4s}$ where the time derivative of the shear stress vanishes (compare Fig.~\ref{fig:pizzprime}).
}
\end{figure}

For small velocity gradients $\frac{1}{\tau T}\ll 1$, corresponding to the Navier-Stokes regime, $\Phi$ and $\Pi$ are well approximated by the first terms in Eq.~(\ref{eq:evol}), respectively. For the shear sector, using the entropy density $s$, one may thus attempt to extract the temperature-dependent ratio $\eta/s$ directly from plotting $\frac{3 \Phi \tau}{4s}$ versus temperature (``Method I'', see Fig.~\ref{fig:eta}). The result thus obtained should be unchanged when initializing the hadron cascade code at later times at the same initial temperature, and we have verified this explicitly by choosing to initialize at $\tau_0=40$ fm/c instead of $\tau_0=10$ fm/c. Before commenting on a second method to extract $\eta/s$ in the hadron phase, let us briefly digress to discuss the bulk stress sector.

In principle, one should be able to extract the temperature dependent $\zeta/s$ value in the hadron gas by considering the estimator $\frac{\Pi \tau}{s}$ as a function of temperature. Unfortunately, to obtain $\Pi$ we require knowledge about the pressure $P$ in the hot hadron gas, which is complicated by the fact that some hadronic species may be in the process of becoming chemically frozen (partial chemical equilibrium, cf. Ref.~\cite{Hirano:2002ds}). Since we lack accurate quantitative knowledge of the chemically non-equilibrium pressure, this introduces a large systematic error in $\Pi$ and hence our estimator for $\zeta/s$. In particular, we have checked explicitly that using the chemically equilibrated pressure of a hadron gas is not adequate to extract accurate information about $\Pi$.
Without detailed quantitative knowledge of the partially chemically equilibrated equation of state applicable to the hadron gas at high temperature, we do not have a viable method to obtain information about $\zeta/s$. Thus we are not able to report on results in the bulk sector in this work, but intend to revisit the bulk viscosity of a hot hadronic gas in a future study.

\begin{figure}[t]
\centerline{\includegraphics[width=0.5\textwidth]{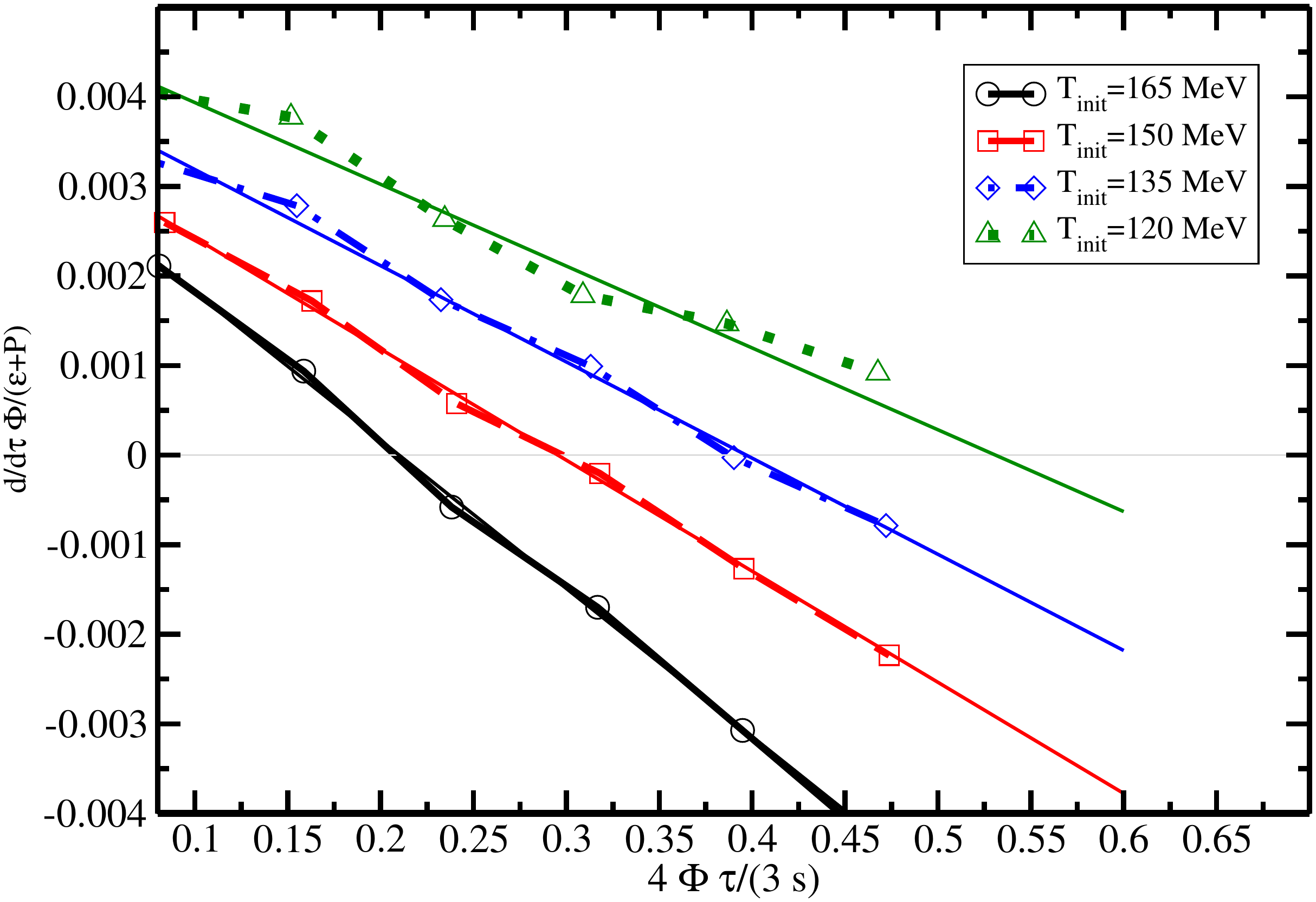}}
\caption{\label{fig:pizzprime} 
Time derivative of $\frac{\Phi}{\epsilon+P}$ at $\tau_0=40$ fm/c as a function of the initial anisotropy $\Phi$, for various initialization temperatures $T_{\rm init}=T(\tau_0)$ (symbols). The values for which the derivative vanishes (partly obtained by linear extrapolation, full lines) point to $\Phi$ being equal to its Navier-Stokes value, which can then be used to determine $\eta/s$ at $T(\tau_0)$.}
\end{figure}

A second method to extract $\eta/s$ is to initialize the system with different initial values for $\Phi$, and should then observe its relaxation towards the Navier-Stokes result. To this end, note that Eq.~(\ref{eq:evol}) can be rewritten as
\begin{eqnarray}
\label{eq:IS}
\partial_\tau \left(\frac{\Phi}{\epsilon+P}\right)&=&- \frac{1}{\tau_\pi}
\left(
\frac{\Phi-\frac{4 \eta}{3\tau}}{\epsilon+P}
\right)\left[1+{\cal O}\left(\frac{1}{\tau T}\right)\right]\,.
%\partial_\tau \left(\frac{\Pi}{s}\right)&=&-\frac{1}{\tau_\Pi}\left(\frac{\Pi-\frac{\zeta}{\tau}}{s}\right)\left[1+{\cal O}\left(\frac{1}{\tau T}\right)\right]\,.
\end{eqnarray}
The left hand side of Eq.~(\ref{eq:IS}) will vanish if the initial value of $\Phi$ will equal the Navier-Stokes value. Once one finds this value, the shear viscosity is then given by the estimator $\frac{3 \Phi \tau}{4s}$. Figure \ref{fig:pizzprime} demonstrates how the Navier-Stokes value of $\Phi$ was determined by plotting the derivative $\partial_\tau \left(\frac{\Phi}{\epsilon+P}\right)$ at $\tau_0=40$ fm/c as extracted from the simulation. The location where the derivative vanishes then graphically points to the viscosity to entropy ratio. Note that Method II is more sensitive to higher order gradient corrections than Method I and hence we have to choose $\tau_0>20$ fm/c to obtain accurate results.
 
Using results from Method I and II, we obtain an estimate for the viscosity of a hot hadron gas with temperatures between 120 MeV and 170 MeV given by
\begin{equation}
\frac{\eta}{s}\simeq 1.45 -1.3 \frac{T}{170\,{\rm MeV}}\pm0.02\,.
\end{equation}
This is our main result.

The extracted viscosity is very much in line with rough expectations from the Kubo formula if one believes that the correlation between deviations of the stress-energy tensor decay in approximately two collision times. However, compared to results in the literature, we find much smaller values for $\eta/s$ than Ref.~\cite{Demir:2008tr}, and somewhat larger values than reported in Ref.~\cite{Song:2010aq}. The discrepancy with respect to Ref.~\cite{Song:2010aq} could be due to the fact that Ref.~\cite{Song:2010aq} initialized their hadron cascade at $T_{SW}=170$ MeV, thus being very sensitive to the transients also seen in Fig.~\ref{fig:eta}. We do not have an explanation for the discrepancy with respect to the results obtained in Ref.~\cite{Demir:2008tr}.

\begin{figure}[h]
\includegraphics[width=0.45\textwidth]{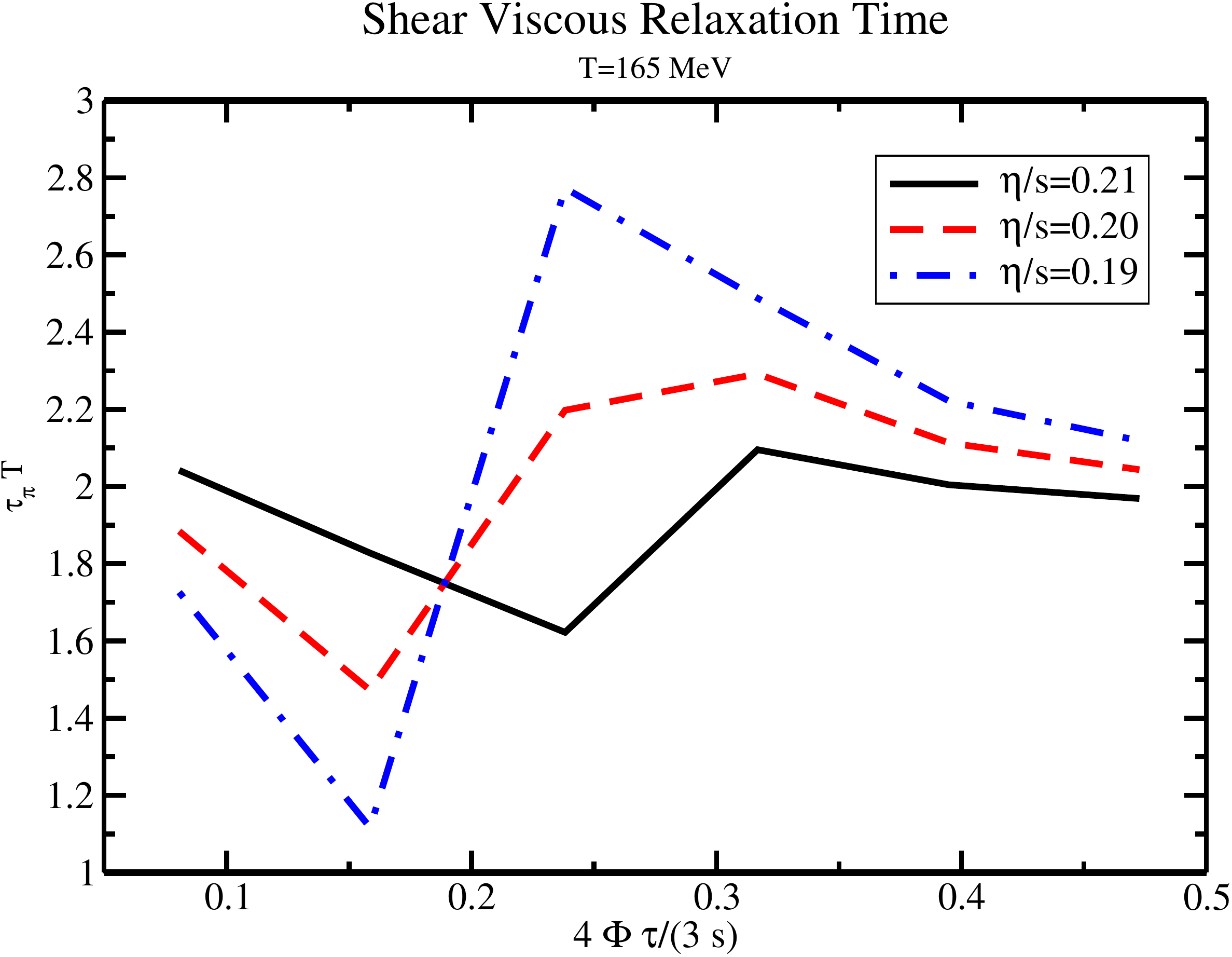}
\caption{\label{fig:taupi}
Shear viscous relaxation time $\tau_\pi$ as a function of shear stress at $T=165$ MeV ($\tau_0=40$ fm/c) from Eq.~(\ref{eq:2o}) for values of $\eta/s$ consistent with those from Fig.~\ref{fig:eta}.}
\end{figure}

Once a value for the shear viscosity over entropy ratio is found, we may also attempt to extract second order transport coefficients such as the shear viscous relaxation time. To this end, we rewrite Eq.~(\ref{eq:IS}) as 
\begin{equation}
\label{eq:2o}
\tau_\pi T \simeq  \frac{\frac{4 \eta}{3\tau s}-\frac{\Phi}{s}}{\partial_\tau \left(\Phi/(\epsilon+P)\right)}\,,
\end{equation}
and plot $\tau_\pi T$ at $T(\tau_0)$ as a function of initialization values for $\Phi$ for various values of $\eta/s$. The result at $T=165$ MeV is shown in Fig.~\ref{fig:taupi}, suggesting that $\tau_\pi T$  is consistent with two, which implies a ratio of $\tau_\pi T s/\eta$ at the high end of values reported in Ref.~\cite{York:2008rr}. 
Note that our result seems consistent with the findings of Ref.~\cite{Muronga:2007qf} for a hot gas of $\pi,\eta,\omega,\rho$ and $\phi$ mesons. 
However, the accuracy of our method needs to be improved in future studies to draw firmer conclusions on the value of second order coefficients in the hot hadron gas.

\section*{Conclusions}

In this work, we have studied transport properties in the hot hadron gas by measuring the energy-momentum tensor in a hadron cascade simulation undergoing longitudinal expansion. We were able to extract a temperature dependent value of the shear viscosity over entropy density for temperature between 120 and 170 MeV. We furthermore investigated the possibility of extracting the bulk viscosity and second-order transport coefficients in the hot hadron gas. While partial chemical non-equilibrium effects seem to prohibit us from extracting $\zeta/s$, we were able to extract an estimate for the shear viscous relaxation time at $T=165$ MeV. Many aspects of our work can be improved in a straightforward manner and we expect our methodology to be useful in quantitative studies of transport in the hot hadron gas in the future.

\begin{acknowledgments}
 
This work was supported by the Department of Energy, awards No. DE-SC0008027 and DE-FG02-03ER41259.

\end{acknowledgments}

\bibliographystyle{apsrev} \bibliography{cascade}

\begin{thebibliography}{22}
\expandafter\ifx\csname natexlab\endcsname\relax\def\natexlab#1{#1}\fi
\expandafter\ifx\csname bibnamefont\endcsname\relax
  \def\bibnamefont#1{#1}\fi
\expandafter\ifx\csname bibfnamefont\endcsname\relax
  \def\bibfnamefont#1{#1}\fi
\expandafter\ifx\csname citenamefont\endcsname\relax
  \def\citenamefont#1{#1}\fi
\expandafter\ifx\csname url\endcsname\relax
  \def\url#1{\texttt{#1}}\fi
\expandafter\ifx\csname urlprefix\endcsname\relax\def\urlprefix{URL }\fi
\providecommand{\bibinfo}[2]{#2}
\providecommand{\eprint}[2][]{\url{#2}}

\bibitem[{\citenamefont{Aoki et~al.}(2006)\citenamefont{Aoki, Endrodi, Fodor,
  Katz, and Szabo}}]{Aoki:2006we}
\bibinfo{author}{\bibfnamefont{Y.}~\bibnamefont{Aoki}},
  \bibinfo{author}{\bibfnamefont{G.}~\bibnamefont{Endrodi}},
  \bibinfo{author}{\bibfnamefont{Z.}~\bibnamefont{Fodor}},
  \bibinfo{author}{\bibfnamefont{S.}~\bibnamefont{Katz}}, \bibnamefont{and}
  \bibinfo{author}{\bibfnamefont{K.}~\bibnamefont{Szabo}},
  \bibinfo{journal}{Nature} \textbf{\bibinfo{volume}{443}},
  \bibinfo{pages}{675} (\bibinfo{year}{2006}), \eprint{hep-lat/0611014}.

\bibitem[{\citenamefont{Bhattacharya et~al.}(2014)\citenamefont{Bhattacharya,
  Buchoff, Christ, Ding, Gupta et~al.}}]{Bhattacharya:2014ara}
\bibinfo{author}{\bibfnamefont{T.}~\bibnamefont{Bhattacharya}},
  \bibinfo{author}{\bibfnamefont{M.~I.} \bibnamefont{Buchoff}},
  \bibinfo{author}{\bibfnamefont{N.~H.} \bibnamefont{Christ}},
  \bibinfo{author}{\bibfnamefont{H.~T.} \bibnamefont{Ding}},
  \bibinfo{author}{\bibfnamefont{R.}~\bibnamefont{Gupta}}, \bibnamefont{et~al.}
  (\bibinfo{year}{2014}), \eprint{1402.5175}.

\bibitem[{\citenamefont{Gavin}(1985)}]{Gavin:1985ph}
\bibinfo{author}{\bibfnamefont{S.}~\bibnamefont{Gavin}},
  \bibinfo{journal}{Nucl.Phys.} \textbf{\bibinfo{volume}{A435}},
  \bibinfo{pages}{826} (\bibinfo{year}{1985}).

\bibitem[{\citenamefont{Prakash et~al.}(1993)\citenamefont{Prakash, Prakash,
  Venugopalan, and Welke}}]{Prakash:1993bt}
\bibinfo{author}{\bibfnamefont{M.}~\bibnamefont{Prakash}},
  \bibinfo{author}{\bibfnamefont{M.}~\bibnamefont{Prakash}},
  \bibinfo{author}{\bibfnamefont{R.}~\bibnamefont{Venugopalan}},
  \bibnamefont{and} \bibinfo{author}{\bibfnamefont{G.}~\bibnamefont{Welke}},
  \bibinfo{journal}{Phys.Rept.} \textbf{\bibinfo{volume}{227}},
  \bibinfo{pages}{321} (\bibinfo{year}{1993}).

\bibitem[{\citenamefont{Dobado and Santalla}(2002)}]{Dobado:2001jf}
\bibinfo{author}{\bibfnamefont{A.}~\bibnamefont{Dobado}} \bibnamefont{and}
  \bibinfo{author}{\bibfnamefont{S.~N.} \bibnamefont{Santalla}},
  \bibinfo{journal}{Phys.Rev.} \textbf{\bibinfo{volume}{D65}},
  \bibinfo{pages}{096011} (\bibinfo{year}{2002}), \eprint{hep-ph/0112299}.

\bibitem[{\citenamefont{Chen and Nakano}(2007)}]{Chen:2006iga}
\bibinfo{author}{\bibfnamefont{J.-W.} \bibnamefont{Chen}} \bibnamefont{and}
  \bibinfo{author}{\bibfnamefont{E.}~\bibnamefont{Nakano}},
  \bibinfo{journal}{Phys.Lett.} \textbf{\bibinfo{volume}{B647}},
  \bibinfo{pages}{371} (\bibinfo{year}{2007}), \eprint{hep-ph/0604138}.

\bibitem[{\citenamefont{Chen et~al.}(2007)\citenamefont{Chen, Li, Liu, and
  Nakano}}]{Chen:2007xe}
\bibinfo{author}{\bibfnamefont{J.-W.} \bibnamefont{Chen}},
  \bibinfo{author}{\bibfnamefont{Y.-H.} \bibnamefont{Li}},
  \bibinfo{author}{\bibfnamefont{Y.-F.} \bibnamefont{Liu}}, \bibnamefont{and}
  \bibinfo{author}{\bibfnamefont{E.}~\bibnamefont{Nakano}},
  \bibinfo{journal}{Phys.Rev.} \textbf{\bibinfo{volume}{D76}},
  \bibinfo{pages}{114011} (\bibinfo{year}{2007}), \eprint{hep-ph/0703230}.

\bibitem[{\citenamefont{Itakura et~al.}(2008)\citenamefont{Itakura, Morimatsu,
  and Otomo}}]{Itakura:2007mx}
\bibinfo{author}{\bibfnamefont{K.}~\bibnamefont{Itakura}},
  \bibinfo{author}{\bibfnamefont{O.}~\bibnamefont{Morimatsu}},
  \bibnamefont{and} \bibinfo{author}{\bibfnamefont{H.}~\bibnamefont{Otomo}},
  \bibinfo{journal}{Phys.Rev.} \textbf{\bibinfo{volume}{D77}},
  \bibinfo{pages}{014014} (\bibinfo{year}{2008}), \eprint{0711.1034}.

\bibitem[{\citenamefont{Bass et~al.}(1998)\citenamefont{Bass, Belkacem,
  Bleicher, Brandstetter, Bravina et~al.}}]{Bass:1998ca}
\bibinfo{author}{\bibfnamefont{S.}~\bibnamefont{Bass}},
  \bibinfo{author}{\bibfnamefont{M.}~\bibnamefont{Belkacem}},
  \bibinfo{author}{\bibfnamefont{M.}~\bibnamefont{Bleicher}},
  \bibinfo{author}{\bibfnamefont{M.}~\bibnamefont{Brandstetter}},
  \bibinfo{author}{\bibfnamefont{L.}~\bibnamefont{Bravina}},
  \bibnamefont{et~al.}, \bibinfo{journal}{Prog.Part.Nucl.Phys.}
  \textbf{\bibinfo{volume}{41}}, \bibinfo{pages}{255} (\bibinfo{year}{1998}),
  \eprint{nucl-th/9803035}.

\bibitem[{\citenamefont{Demir and Bass}(2009)}]{Demir:2008tr}
\bibinfo{author}{\bibfnamefont{N.}~\bibnamefont{Demir}} \bibnamefont{and}
  \bibinfo{author}{\bibfnamefont{S.~A.} \bibnamefont{Bass}},
  \bibinfo{journal}{Phys.Rev.Lett.} \textbf{\bibinfo{volume}{102}},
  \bibinfo{pages}{172302} (\bibinfo{year}{2009}), \eprint{0812.2422}.

\bibitem[{\citenamefont{Song et~al.}(2011)\citenamefont{Song, Bass, and
  Heinz}}]{Song:2010aq}
\bibinfo{author}{\bibfnamefont{H.}~\bibnamefont{Song}},
  \bibinfo{author}{\bibfnamefont{S.~A.} \bibnamefont{Bass}}, \bibnamefont{and}
  \bibinfo{author}{\bibfnamefont{U.}~\bibnamefont{Heinz}},
  \bibinfo{journal}{Phys.Rev.} \textbf{\bibinfo{volume}{C83}},
  \bibinfo{pages}{024912} (\bibinfo{year}{2011}), \eprint{1012.0555}.

\bibitem[{\citenamefont{Bjorken}(1983)}]{Bjorken:1982qr}
\bibinfo{author}{\bibfnamefont{J.}~\bibnamefont{Bjorken}},
  \bibinfo{journal}{Phys.Rev.} \textbf{\bibinfo{volume}{D27}},
  \bibinfo{pages}{140} (\bibinfo{year}{1983}).

\bibitem[{\citenamefont{Baier et~al.}(2008)\citenamefont{Baier, Romatschke,
  Son, Starinets, and Stephanov}}]{Baier:2007ix}
\bibinfo{author}{\bibfnamefont{R.}~\bibnamefont{Baier}},
  \bibinfo{author}{\bibfnamefont{P.}~\bibnamefont{Romatschke}},
  \bibinfo{author}{\bibfnamefont{D.~T.} \bibnamefont{Son}},
  \bibinfo{author}{\bibfnamefont{A.~O.} \bibnamefont{Starinets}},
  \bibnamefont{and} \bibinfo{author}{\bibfnamefont{M.~A.}
  \bibnamefont{Stephanov}}, \bibinfo{journal}{JHEP}
  \textbf{\bibinfo{volume}{0804}}, \bibinfo{pages}{100} (\bibinfo{year}{2008}),
  \eprint{0712.2451}.

\bibitem[{\citenamefont{Romatschke}(2010)}]{Romatschke:2009kr}
\bibinfo{author}{\bibfnamefont{P.}~\bibnamefont{Romatschke}},
  \bibinfo{journal}{Class.Quant.Grav.} \textbf{\bibinfo{volume}{27}},
  \bibinfo{pages}{025006} (\bibinfo{year}{2010}), \eprint{0906.4787}.

\bibitem[{\citenamefont{Bhattacharyya et~al.}(2008)\citenamefont{Bhattacharyya,
  Hubeny, Minwalla, and Rangamani}}]{Bhattacharyya:2008jc}
\bibinfo{author}{\bibfnamefont{S.}~\bibnamefont{Bhattacharyya}},
  \bibinfo{author}{\bibfnamefont{V.~E.} \bibnamefont{Hubeny}},
  \bibinfo{author}{\bibfnamefont{S.}~\bibnamefont{Minwalla}}, \bibnamefont{and}
  \bibinfo{author}{\bibfnamefont{M.}~\bibnamefont{Rangamani}},
  \bibinfo{journal}{JHEP} \textbf{\bibinfo{volume}{0802}}, \bibinfo{pages}{045}
  (\bibinfo{year}{2008}), \eprint{0712.2456}.

\bibitem[{\citenamefont{Denicol et~al.}(2012)\citenamefont{Denicol, Molnár,
  Niemi, and Rischke}}]{Denicol:2012es}
\bibinfo{author}{\bibfnamefont{G.}~\bibnamefont{Denicol}},
  \bibinfo{author}{\bibfnamefont{E.}~\bibnamefont{Molnár}},
  \bibinfo{author}{\bibfnamefont{H.}~\bibnamefont{Niemi}}, \bibnamefont{and}
  \bibinfo{author}{\bibfnamefont{D.}~\bibnamefont{Rischke}},
  \bibinfo{journal}{Eur.Phys.J.} \textbf{\bibinfo{volume}{A48}},
  \bibinfo{pages}{170} (\bibinfo{year}{2012}), \eprint{1206.1554}.

\bibitem[{\citenamefont{Muronga}(2004)}]{Muronga:2003ta}
\bibinfo{author}{\bibfnamefont{A.}~\bibnamefont{Muronga}},
  \bibinfo{journal}{Phys.Rev.} \textbf{\bibinfo{volume}{C69}},
  \bibinfo{pages}{034903} (\bibinfo{year}{2004}), \eprint{nucl-th/0309055}.

\bibitem[{\citenamefont{Novak et~al.}(2013)\citenamefont{Novak, Novak, Pratt,
  Coleman-Smith, and Wolpert}}]{Novak:2013bqa}
\bibinfo{author}{\bibfnamefont{J.}~\bibnamefont{Novak}},
  \bibinfo{author}{\bibfnamefont{K.}~\bibnamefont{Novak}},
  \bibinfo{author}{\bibfnamefont{S.}~\bibnamefont{Pratt}},
  \bibinfo{author}{\bibfnamefont{C.}~\bibnamefont{Coleman-Smith}},
  \bibnamefont{and} \bibinfo{author}{\bibfnamefont{R.}~\bibnamefont{Wolpert}}
  (\bibinfo{year}{2013}), \eprint{1303.5769}.

\bibitem[{\citenamefont{Pratt and Torrieri}(2010)}]{Pratt:2010jt}
\bibinfo{author}{\bibfnamefont{S.}~\bibnamefont{Pratt}} \bibnamefont{and}
  \bibinfo{author}{\bibfnamefont{G.}~\bibnamefont{Torrieri}},
  \bibinfo{journal}{Phys.Rev.} \textbf{\bibinfo{volume}{C82}},
  \bibinfo{pages}{044901} (\bibinfo{year}{2010}), \eprint{1003.0413}.

\bibitem[{\citenamefont{Hirano and Tsuda}(2002)}]{Hirano:2002ds}
\bibinfo{author}{\bibfnamefont{T.}~\bibnamefont{Hirano}} \bibnamefont{and}
  \bibinfo{author}{\bibfnamefont{K.}~\bibnamefont{Tsuda}},
  \bibinfo{journal}{Phys.Rev.} \textbf{\bibinfo{volume}{C66}},
  \bibinfo{pages}{054905} (\bibinfo{year}{2002}), \eprint{nucl-th/0205043}.

\bibitem[{\citenamefont{York and Moore}(2009)}]{York:2008rr}
\bibinfo{author}{\bibfnamefont{M.~A.} \bibnamefont{York}} \bibnamefont{and}
  \bibinfo{author}{\bibfnamefont{G.~D.} \bibnamefont{Moore}},
  \bibinfo{journal}{Phys.Rev.} \textbf{\bibinfo{volume}{D79}},
  \bibinfo{pages}{054011} (\bibinfo{year}{2009}), \eprint{0811.0729}.

\bibitem[{\citenamefont{Muronga}(2008)}]{Muronga:2007qf}
\bibinfo{author}{\bibfnamefont{A.}~\bibnamefont{Muronga}},
  \bibinfo{journal}{Eur.Phys.J.ST} \textbf{\bibinfo{volume}{155}},
  \bibinfo{pages}{107} (\bibinfo{year}{2008}), \eprint{0710.3280}.

\end{thebibliography}

\end{document}